\documentclass[10pt]{article} 
\usepackage{cite}
\usepackage{latexsym}
\usepackage{amsmath}
\usepackage{indentfirst}
\usepackage{pstricks}

\long\def\ca#1\cb{} 

\newcommand{\AND}{{\small AND}}

\newcommand{\becs}{\begin{cases}}
\newcommand{\bem}{\begin{matrix}}

\newcommand{\dya}[1]{|#1\rangle\langle#1|}

\newcommand{\encs}{\end{cases}}
\newcommand{\enm}{\end{matrix}}

\newcommand{\ket}[1]{|#1\rangle }

\newcommand{\NOT}{{\small NOT}}

\newcommand{\OR}{{\small OR}}

 \def\outl#1{}  \def\xa{} \def\xb{}  

\ca
 \def\outl#1{\par{\medskip\noindent\hspace*{.5cm}\bf
      \mathversion{bold}#1\mathversion{normal}\smallskip} }
 \def\xa{} \def\xb{}  
\cb

\ca
 \def\outl#1{\par{\medskip\noindent\hspace*{.5cm}\bf
      \mathversion{bold}#1\mathversion{normal}\smallskip} }
 \long\def\xa#1\xb{}
 
\cb

\begin{document}

\title{The Logic of Consistent Histories: A Reply to Maudlin}

\author{Robert B. Griffiths%
\thanks{Electronic mail: rgrif@cmu.edu}\\ 
Department of Physics,
Carnegie-Mellon University,\\
Pittsburgh, PA 15213, USA}

\date{
Version of 26 September 2011}

\maketitle  

\xa
\begin{abstract}

  The relationship between quantum logic, standard propositional logic, and
  the (consistent) histories rules for quantum reasoning is discussed.  It is
  shown that Maudlin's claim \cite{Mdln11}, that the histories approach is
  inconsistent, is incorrect.  The histories approach is both internally
  consistent and adequate for discussing the physical situations considered by
  Maudlin. 

\end{abstract}
\xb

\tableofcontents
\xa

\xb
\section{Introduction}
\label{sct1}
\xa

\xb
\outl{Maudlin's inconsistency claim is wrong}
\xa

The argument for the locality of quantum mechanics in \cite{Grff11b} based on
the (``consistent'' or ``decoherent'') histories interpretation has been
criticized by Maudlin \cite{Mdln11} who claims that the histories approach is
inconsistent.  It will be shown here that Maudlin's arguments are incorrect:
the histories approach is both internally consistent as a form of quantum
logic, and also an appropriate tool for consistently analyzing the two specific
physical situations mentioned in \cite{Mdln11}.
Naturally, a system of reasoning that is internally consistent may not be
appropriate or adequate when applied to a particular problem, and it could
be that Maudlin's complaint about inconsistency is really about, or perhaps
includes, something which might better be called inadequacy.  In what
follows some attempt will be made to address adequacy along with consistency
in the narrower sense, though the emphasis will be on the latter. 

In Sec.~\ref{sct2} a brief comparison is made between (classical)
propositional (or sentential) logic, the quantum logic first proposed by
Birkhoff and von Neumann \cite{BrvN36}, and the histories approach, with
particular attention on the consistency of these three systems.  Following
that, Secs.~\ref{sct3} and \ref{sct4} address two physical situations
mentioned in \cite{Mdln11}: spins prepared and placed in labeled boxes, and,
briefly, the GHZ paradox.  Section~\ref{sct5} contains a short summary.

\xb
\section{Logic and Quantum Mechanics}
\label{sct2}
\xa

\xb
\subsection{Propositional logic}
\label{sbct2.1}
\xa

In discussing the consistency of the histories approach to quantum mechanics
it is helpful to first review a few features of formal logic. We need only
consider propositional (sentential) logic,%
\footnote{This branch of logic is discussed in many books. A readable (albeit
  longwinded) internet resource will be found at \cite{GrOs04}.}
not the more complicated (first-order) predicate logic.  The
rules of the former are of two types.  First come rules for putting together
formulas (also called sentences or statements or propositions) in a meaningful
way using logical connectives.  Thus $\lnot A$ (read as ``\NOT\ $A$''),
$A\land B$ (``$A$ \AND\ $B$''), and $A\lor B$ (``$A$ \OR\ $B$'') are allowed
by these rules, provided $A$ and $B$ (think of them as assertions which could
be true or false) are themselves propositions or symbols in an approved list.
A combination such as $A\land\lor B$ is not allowed; it is not
meaningful. Sometimes one refers to expressions which have been properly
constructed as ``well-formed formulas.''  Second come rules that govern
logical inference: the validity of an argument which begins with a set
\emph{premises}, each of which must itself be a well-formed formula, and ends
in a \emph{conclusion}, also a well-formed formula.  The basic idea behind the
rules is to make sure that the truth of the conclusion follows from the truth
of (at least some of) the premises, based solely on the logical structure of
the argument; i.e., that the argument is \emph{valid}. Logicians are not in
the business of guaranteeing the truth of the premises that enter an argument;
instead their goal is identify valid arguments.  For example, from the truth
of $A\land B$ as a premise it follows as a conclusion that $A$ is true, but
this is not so if the premise is, instead, $A\lor B$.

\xb
\outl{Consistency of propositional logic}
\xa

A system of propositional logic is considered \emph{inconsistent} if its rules
allow one to reason to the truth of both a proposition and its negation, to
the truth of $A$ and of $\lnot A$, from the
same premises.  Standard propositional logic is known to be consistent
provided one follows the rules worked out for it by logicians. The consistency
of alternative logics must, of course, be assessed on the basis of their own
rules.

\xb
\subsection{Quantum Logic}
\label{sbct2.2}
\xa

\xb
\outl{Introduction to Qm logic using subspaces}
\xa

Quantum logic was discussed briefly in \cite{Grff11b} and \cite{Mdln11}; see
\cite{AdWr83} for an accessible treatment with more details.  Propositions
such as $A$ and $B$ correspond to subspaces of a complex Hilbert space, which
for present purposes can be considered finite dimensional.  Negation $\lnot A$
corresponds to the orthogonal complement of a subspace, $A\land B$ to the
intersection $A\cap B$ of two subspaces, and $A\lor B$ to their direct sum
$A\oplus B$.  The resulting logic resembles but is not identical to standard
propositional logic; in particular the distributive laws of the latter no
longer hold for quantum logic.  This has as a consequence the fact that if one
insists on applying the rules of (standard) propositional logic to the
formulas of quantum logic it is possible to derive contradictory results.
See, for example, the discussion of Eq.~(2) in \cite{Grff11b}. This does
\emph{not} imply that quantum logic is inconsistent; its consistency must be
checked using its own rules, and assessed by these rules it provides a
consistent system of reasoning \cite{Mttl09}.

\xb
\outl{$A\land B$ and $A\lor B$ in quantum logic and propositional logic}
\xa

It is worth noting that whereas the connectives $\land$ and $\lor$ are defined
differently in quantum and in ordinary propositional logic, the two are
closely connected in the following sense.  If the projectors $A$ and $B$ for
two subspaces commute, $AB=BA$, then (using the same symbols for a space and
its projector) $A\land B$ and $A\lor B$ behave very much like their standard
counterparts, both formally and intuitively, when understood as representing
quantum properties.  Indeed, if one starts with any set of \emph{commuting}
projectors and repeatedly applies the quantum logical connectives (including
$\lnot$) to these projectors and to those that result from earlier
applications, the result is a collection of projectors that continue to
commute with each other and form what is called a Boolean algebra (or Boolean
lattice). All the ideas of propositional logic then apply without any change
to this ``commuting fragment'' (as we might call it) of quantum logic.  In
brief, as long as one fixes one's attention on a collection of commuting
projectors, quantum logic is the same as standard propositional logic;
differences only appear when projectors fail to commute.  And since the
physical properties considered in classical mechanics correspond, when viewed
quantum mechanically, as a collection of commuting projectors, see Sec.~26.6
of \cite{Grff02c}, there is, contrary to \cite{Mdln11}, a very good reason to
refer to ordinary propositional logic as ``classical logic'' when
distinguishing it from alternatives used in quantum theory.

\xb
\subsection{Histories Logic and Its Consistency}
\label{sbct2.3}
\xa

\xb
\outl{CH logic differs from Qm logic in the single framework rule}
\xa

The logic used in histories quantum mechanics, as explained in Sec.~II of
\cite{Grff11b}, employs subspaces (equivalently, their projectors) and
negation $\lnot$ in exactly the same way as in quantum logic.  The crucial
difference between the two is the \emph{single framework rule} of the
histories approach.  A \emph{framework} consists of a set of \emph{commuting}
projectors that form a Boolean algebra. Within such a framework the
connectives are the same as those used in quantum logic, and as noted above,
correspond exactly to those in ordinary propositional logic as it is employed
in classical physics.  The single framework rule states that in forming
propositions only projectors (equivalently, quantum properties or subspaces)
belonging to a fixed framework may be employed, and in constructing a logical
argument all the premises and also the conclusion must belong to the same
framework.  This rule is best thought of as analogous to the rules of
propositional logic that define meaningful propositions; what it excludes
are the quantum counterparts of things like $A\land\lor B$.

As a consequence of the single framework rule, in all situations in which
histories logic permits the use of connectives their meaning coincides with
the same symbols whether used in quantum or in (classical) propositional
logic, a point overlooked in \cite{Mdln11}. Its logical rules, always applied
within the given framework (but remember that combinations using different
frameworks are not allowed by these rules) thus coincide with those of quantum
logic as applied in this limited context, and also with the usual rules of
standard propositional logic.
Consequently, a properly constructed argument that follows the rules of
histories quantum mechanics, and is thus necessarily restricted to a single
framework, has precisely the same structure as an argument in standard
propositional logic, and can never lead to a contradiction. Any proof of the
consistency of propositional logic applies equally to the logic used in
consistent histories. In particular, Maudlin's arguments in \cite{Mdln11} for
the consistency of classical logic imply the consistency of the histories
approach, contradicting his claim that the latter is inconsistent.

\xb \outl{Single framework rule allows different frameworks; forbids combining
  them} \xa

Note that, as discussed in more detail in \cite{Grff11b}, the single framework
rule is not at all a prohibition against the physicist constructing various
distinct and mutually incompatible frameworks in the course of analyzing some
quantum mechanical problem. What the rule forbids is \emph{combining}
conclusions from \emph{different} frameworks. It is this that prevents the
inference of logical contradictions, and ensures the consistency of this kind
of quantum reasoning.
For more details, including the consistency of arguments involving time
development (for which there are additional complications, though the basic
principles are the same) see Ch.~16 of \cite{Grff02c}.

\xb
\subsection{Role of Logic in Quantum Theory}
\xa

\xb
\outl{Proponents think Qm logic needed in specifically Qm circumstances}
\xa

Proponents of quantum logic and proponents of the histories approach do not
think of these as replacements for standard propositional logic in terms of
everyday reasoning, or the logic required to write scientific papers, both of
which are in any case only partially represented in propositional logic.
Instead, the idea is that when dealing with specifically quantum phenomena one
needs a mode of reasoning for a situation for which ordinary logic is
inadequate.  With reference to quantum logic see the very helpful discussion
at the beginning of \cite{Bccg09}.  Of course, if the whole world is quantum
mechanical it is reasonable to expect propositional logic as applied in
classical physics to emerge from an appropriate quantum logic as an adequate
approximation to the latter in those situations in which classical mechanics
provides a good approximation to quantum mechanics.  And there is a sound
basis for this expectation in the fact that, as noted above, both quantum and
histories logic are identical with propositional logic, formally and in terms
of their intuitive contents, in physical situations where one is only
interested in a Boolean subalgebra of quantum projectors.

Thus both quantum logic and histories logic are best viewed as extensions of
ordinary propositional logic to a domain in which quantum phenomena are
important, where one might plausibly expect that modifications of the
traditional rules of reasoning are required.  Just as, to use an analogy, the
rules of ordinary arithmetic for products of numbers, where the order does not
matter, have to be modified if one is considering noncommuting operators.
This point seems to have been overlooked by Maudlin in \cite{Mdln11}.  What he
refers to as ``Bell's logic'' can be impeccable but still fail to apply to
something in the quantum domain if for the latter a different mode of
reasoning is more appropriate.  The issue is whether such an alternative
system of reasoning can clear up the the well-known conceptual difficulties of
quantum theory, not whether its rules agree with propositional logic.  Bell
was himself well aware \cite{Bll90} that he had not solved the measurement
problem, and a similar humility might be appropriate on the part of others who
are in the same situation.

We are now in a position to respond to objections by Maudlin \cite{Mdln11} to
the histories approach framed in terms of two specific physical situations.

\xb
\section{Spins Prepared in Boxes}
\label{sct3}
\xa According to the histories approach the question ``Is $S_x=+1/2$ or is
$S_z=+1/2$?''  referring to a particular spin-half particle at a particular
moment in time is meaningless because the projectors, which we shall denote by
$[x^+]=\dya{x^+}$ and $[z^+]$, referring these two possibilities do not
commute.%
\footnote{Maudlin at the end of Sec.~II of \cite{Mdln11} incorrectly replaces
  noncommuting projectors with the condition that the projectors have no
  common eigenstate.  The two are not equivalent in a Hilbert space of
  dimension 3 or more.  However, the distinction is not critical for the
  following discussion.}
Maudlin claims that this represents an inconsistency in histories quantum
mechanics. Imagine, he says, spin half particles prepared either in the state
$S_x=+1/2$ or in the state $S_z=+1/2$, and placed in boxes with each box
carrying a label indicating the preparation procedure.  If a label falls off a
box it is, he asserts, obviously meaningful to ask the question whether the
particle in this particular box is in the state $S_x=+1/2$ or $S_z=+1/2$.

To see what is problematical about this assertion, ask the question: ``How
might we distinguish \emph{by means of a measurement} whether the particle in
this box is in the state $S_x=+1/2$ or in the state $S_z=+1/2$?''  All
students of quantum mechanics know that no measurement can give a definite
answer to this question.  If, for example a measurement of $S_z$ is carried
out by means of a Stern-Gerlach apparatus and the result is $+1/2$, this may
be because the particle was prepared in the state $S_z=+1/2$, but even if it
was prepared in the state $S_x=+1/2$ there is a probability of 1/2 that an
$S_z$ measurement will yield $+1/2$ rather than $-1/2$.%
\footnote{While no measurement on a \emph{single} particle can reliably
  distinguish $S_z=+1/2$ from $S_x=+1/2$, the situation is different if a
  large number $N$ of particles are either (i) all prepared in the same state
  $S_z=+1/2$, or (ii) all prepared in the same state $S_x=+1/2$.  When $N$ is
  large these two situations can be reliably distinguished, i.e., with little
  probability of error, because the quantum states corresponding to (i) and
  (ii) in the total tensor product Hilbert space of dimension $2^N$ are nearly
  orthogonal, and thus distinguishable, even though for a single particle the
  states $\ket{x^+}$ and $\ket{z^+}$ are very far from being orthogonal}
\ %

One does not have to be a logical positivist to be suspicious about a
supposedly obvious distinction which cannot be subjected to experimental test.
Should we perhaps distinguish between a particle \emph{being} in a state with
some definite value of some component of spin, and its being \emph{prepared}
in such a state, even in a situation in which there is no magnetic field
acting to cause spin precession after the measurement?  Indeed we should, and
the histories approach provides the tools needed to make this distinction,
through its concept of a \emph{dependent} or \emph{contextual} state, Ch.~14
of \cite{Grff02c}.  Suppose the particle was prepared in the state $[x^+]$,
$S_x=+1/2$, and a record, Maudlin's label, was made. We can represent the
situation by a quantum projector $[x^+]\otimes L_x$, where $L_x$ projects on
the quantum state of the label.  Similarly $[z^+]\otimes L_z$ is the state of
the particle-plus-label corresponding to a preparation of $S_z = +1/2$.  Since
$L_x$ and $L_z$ refer to distinct macroscopic states, $L_x L_z=0$ (at least to
an excellent approximation). Consequently, $[x^+]\otimes L_x$ and
$[z^+]\otimes L_z$ are also orthogonal.  Thus in the histories approach
``$[x^+]\otimes L_x$ \OR\ $[z^+]\otimes L_z$'' makes sense, even though
``$[x^+]$ \OR\ $[z^+]$'' does not.

Thus quantum mechanics itself, even in the inadequate presentation found in
current textbooks, with measurements an unanalyzed black box, undermines
Maudlin's claim that his spin-half example demonstrates something wrong with
the histories approach. And it is precisely the histories approach that
provides the sorts of distinctions, along with the logical tools needed to
make sense of them, required for a fully quantum-mechanical analysis of
preparations, measurements, and microscopic quantum states.%
\footnote{ For more on these topics, see \cite{Grff11c}.}

\section{GHZ Paradox}
\label{sct4}

For an accessible account of the Greenberger-Horne-Zeilinger or GHZ paradox
\cite{GrHZ89} see the article by Mermin \cite{Mrmn90} that appeared shortly
afterwards, or his later review paper \cite{Mrmn93}.  As discussed in
\cite{Mdln11}, it involves three spin-half particles in a particular quantum
state, at three separate locations so that they do not interact with each
other, each of which can be subjected to two types of measurement, of either
$S_x$ or $S_y$.  The measurement apparatuses and the means of deciding which
spin component is to be measured are located close to the respective particles
and sufficiently far apart (e.g., at spacelike separation in the relativistic
sense) that they cannot influence each other. If one assumes that each
measurement reveals a pre-existing property, e.g., $S_y=+1/2$ for particle 2,
then it is impossible to assign simultaneous $S_x$ and $S_y$ values to each
particle in such a way that the measurement statistics will agree with those
predicted by quantum theory for the specified initial quantum state.  This is
the paradox.  Maudlin claims that any local theory must arrive at this
paradox, and hence a theory which agrees with the predictions of quantum
mechanics for the statistics of measurement outcomes is necessarily nonlocal.

The histories approach does not allow for simultaneous assignment of $S_x$ and
$S_y$ values to a single particle, and thus blocks this route to a paradox by
applying its single framework rule.  Maudlin acknowledges this when he asserts
(Sec.~V) that ``the single framework rule is supposed to prevent us from
making this calculation,'' by which he means the calculation that leads to the
paradox.  Indeed, the single framework rule does precisely that, and in this
respect the histories approach is perfectly consistent according to its own
rules.  Maudlin has not located any inconsistency in the histories
approach. Instead he is pointing out that its rules are different from those
he wishes to apply to the problem at hand.  His claim that the histories
approach is inconsistent is like claiming that the use of Riemannian geometry
in general relativity is inconsistent because its rules are different from
those of Euclidean geometry.%
\footnote{This analogy was used by Putnam\cite{Ptnm68b} when he was advocating
  the use of quantum logic.}

Many other issues could be discussed with reference to the GHZ paradox, such
as the relationship of measurement outcomes, represented in proper quantum
mechanical fashion, to prior states of the measured particles.  They require
the use of specific frameworks, and since there are a large number of
possibilities their analysis lies outside the scope of the present paper.  The
interested reader may wish to look at the very detailed analysis, using the
histories approach, of Hardy's paradox \cite{Hrdy92}, which poses essentially
the same conceptual issues (and has led to similar claims of nonlocality) as
does the GHZ paradox, in Ch.~25 of \cite{Grff02c}.

\xb
\section{Conclusion}
\label{sct5}
\xa

Maudlin's claim in \cite{Mdln11}, that the histories approach discussed in
\cite{Grff11b} is inconsistent, is incorrect.  He has not located any
inconsistency in the logical rules employed in the histories approach.
Instead, his own arguments for the consistency of propositional logic imply
the consistency of the logic of histories, as explained in Sec.~\ref{sbct2.3}.
What he appears to find objectionable is that the histories rules are not
identical to those of classical propositional logic.  This difference,
however, is not grounds for declaring the histories rules inconsistent.  The
rules of the older quantum logic also differ from those of classical logic,
something to which Maudlin does not seem to object.  The consistency of each
system needs to be evaluated by its own rules, not by whether it can be
combined with something else.  Naively mixing the rules of quantum logic with
those of classical logic can lead to inconsistencies, and the same is true if
one mixes the histories rules with the classical rules.

In support of his claim for the inconsistency of the histories approach
Maudlin uses two examples: spins in boxes, and the GHZ paradox.  His analysis
of the former is flawed, for he does not properly distinguish the preparation
of quantum states from their measurement.  It is the histories approach that
provides the precise analysis needed to make sense of this situation.  In the
case of GHZ, Maudlin correctly notes that the histories approach blocks the
attempt to construct a paradox.  In doing so it (once again) pulls the rug out
from under an attempt to demonstrate nonlocality in the quantum world.  But
that does not indicate any inconsistency in the histories analysis.

The main problem with quantum logic is not any internal inconsistency, or that
it leads to results that contradict the empirically well-established formulas
of textbook quantum theory.  Instead, it is that this approach has not managed
to resolve the conceptual difficulties of quantum mechanics, something even
even its proponents admit; for a recent assessment see \cite{Bccg09}.

The histories approach is also internally consistent, and agrees with all
experiments that confirm the validity of quantum theory. It provides a fully
quantum-mechanical description of how physical measurements actually work,
thus resolving the measurement problem.  In addition it can handle numerous
paradoxes, see Chs.~20 to 25 of \cite{Grff02c}, that cause difficulties for
other interpretations of quantum mechanics.  Its success in resolving these
problems suggests that it deserves much more serious study on the part of
physicists and philosophers interested in quantum foundations than it has
hitherto received \cite{Hhnb10}.  Needless to say, such study must begin with
a careful effort to understand the rules of reasoning used in the histories
approach and how these are applied to various examples.  There may be
fundamental inconsistencies in the histories approach, and the community would
benefit from having them pointed out.  However, those who suspect that
something is amiss might wish to first pay attention to previous claims of
this sort, and the manner in which they have been refuted; see Sec.~8.2 of
\cite{Grff11c} for references.

\section*{Acknowledgments}

I am indebted to T. Maudlin for comments on a draft of this
manuscript. Support for this research has come from the National Science
Foundation through Grants PHY-0757251 and PHY-1068331.

\xb

\end{document}